\documentclass[aps,prb,amsmath,amssymb,twocolumn,showpacs,floatfix]{revtex4-1}
\usepackage{graphicx}% Include figure files
\usepackage{dcolumn}% Align table columns on decimal point
\usepackage{bm}% bold math

\begin{document}
%%%%%%%%%%%%%%%%%%%%%%%%%%%%%%%%%%%%%%%%%%%%%%%%%%%
\title{Investigation of potential fluctuating intra-unit cell magnetic order in cuprates by $\mu$SR}

\author{A.~Pal,$^1$ K.~Akintola,$^1$ M.~Potma,$^1$ M.~Ishikado,$^2$ H.~Eisaki,$^3$ W.N.~Hardy,$^{4,5}$ D.A.~Bonn,$^{4,5}$ R.~Liang,$^{4,5}$ and J.E.~Sonier,$^{1,5}$}

\affiliation{$^1$Department of Physics, Simon Fraser University, Burnaby, British Columbia V5A 1S6, Canada \\
$^2$Research Center for Neutron Science and Technology, Tokai, Naka, Ibaraki, Japan 319-1106 \\
$^3$National Institute of Advanced Industrial Science and Technology, Tsukuba, Ibaraki, Japan 305-8568 \\
$^4$Department of Physics and Astronomy, University of British Columbia,
Vancouver, British Columbia V6T 1Z1, Canada \\
$^5$Canadian Institute for Advanced Research, Toronto, Ontario M5G 1Z8, Canada}

\date{\today}
%%%%%%%%%%%%%%%%%%%%%%%%%%%%%%%%%%%%%%%%%%%%%%%%%%%%%%%
\begin{abstract}
We report low temperature muon spin relaxation ($\mu$SR) measurements of the high-transition-temperature ($T_c$) 
cuprate superconductors Bi$_{2+x}$Sr$_{2-x}$CaCu$_2$O$_{8+\delta}$ and YBa$_2$Cu$_3$O$_{6.57}$, aimed at detecting the mysterious 
intra-unit cell (IUC) magnetic order that has been observed by spin polarized neutron scattering in the pseudogap phase of four different cuprate families. 
A lack of confirmation by local magnetic probe methods has raised the possibility that the magnetic order fluctuates slowly enough to appear static 
on the time scale of neutron scattering, but too fast to affect $\mu$SR or nuclear magnetic resonance (NMR) signals. 
The IUC magnetic order has been linked to a theoretical model for the cuprates, which predicts a long-range 
ordered phase of electron-current loop order that terminates at a
quantum crictical point (QCP). Our study suggests that lowering the temperature to $T \! \sim \! 25$~mK and moving far below the purported QCP
does not cause enough of a slowing down of fluctuations for the IUC magnetic order to become detectable on the time scale of $\mu$SR.
Our measurements place narrow limits on the fluctuation rate of this unidentified magnetic order.
\end{abstract}
\maketitle
%%%%%%%%%%%%%%%%%%%%%%%%%%%%%%%%%%%%%%%%%%%%%%%

An enduring and central open question concerning cuprate superconductors is the nature of the mysterious pseudogap regime above $T_c$. Achieving an understanding 
of the pseudogap (PG) has long been viewed as key to understanding high-$T_c$ superconductivity. A clue to the origin of the PG has come from spin-polarized 
neutron diffraction studies that have detected the onset of an unusual three-dimensional (3-D), long-range IUC magnetic order at a temperature 
concomitant with the PG onset temperature $T^*$ in YBa$_2$Cu$_3$O$_{6+x}$ (Y123), HgBa$_2$CuO$_{4+\delta}$ (Hg1201) and 
Bi$_2$Sr$_2$CaCu$_2$O$_{8+\delta}$ (Bi2212).\cite{Fauque:06,Mook:08,Li:08,Baledent:11,Li:11,Didry:12,Thro:14} This finding provides evidence for a change 
in symmetry at $T^*$ associated with the onset of a novel type of order, which is supported by other kinds of measurements that indicate the
the PG is related to a true phase transition.\cite{Xia:08,Leridon:09,He:11,Shekhter:13}
The magnetic order observed by polarized neutron diffraction is described by staggered out-of-plane magnetic moments that diminish in
magnitude from the underdoped to optimally-doped regime.\cite{Thro:14,Thro:15} 
A similar mysterious magnetic order is also observed in $x \! = \! 0.085$ La$_{2-x}$Sr$_x$CuO$_4$ (LSCO),\cite{Baledent:10} 
although it is short-range, two-dimensional, and onsets at a temperature far below $T^*$. The latter is also the case in underdoped YBa$_2$Cu$_3$O$_{6.45}$ 
--- suggesting a potential competition with Cu spin density wave order at low doping. 

The magnetic structure and the hole-doping dependence of the onset temperature of the IUC magnetic order are somewhat compatible with a 
model derived from a three-band Hubbard model, which attributes the PG to a time-reversal symmetry breaking phase consisting of a pattern of
circulating electron currents that preserve translational symmetry.\cite{Varma:97} With increased hole doping the transition temperature of the 
circulating-current (CC) ordered phase is reduced towards zero, terminating at a QCP within the superconducting phase near
or above optimal doping. Yet zero-field (ZF) $\mu$SR experiments have found no evidence for such a 
magnetically ordered phase.\cite{Sonier:02,MacDougall:08,Sonier:09,Huang:12} 
While it has been suggested that charge screening of the positively charged muon ($\mu^+$) causes severe underdoping of its local 
environment, resulting in the loss of CC order over a distance of several lattice constants,\cite{Shekhter:08} such severe 
perturbation of the local environment is inconsistent with $\mu^+$-Knight shift 
measurements that show a linear scaling with the bulk magnetic susceptibility.\cite{Kaiser:12} Moreover, non-perturbative NMR and nuclear quadupole resonance (NQR)
experiments also find no evidence of IUC magnetic order.\cite{Strassle:08,Strassle:10,Mounce:13,Wu:15} 
It has been argued from calculations in a multi-orbital Hubbard model and for parameters relevant to cuprate superconductors, that the CC phase 
proposed in Ref.~\onlinecite{Varma:97} or variations of it are unlikely to be stabilized as the ground state.\cite{Kung:14} 
A staggered ordering of Ising-like oxygen orbital magnetic moments has been offered as an alternative explanation of the
IUC magnetic order.\cite{Moskvin:12}

Since the original CC phase proposal, the model has been extended to include quantum critical fluctuations of the CC order parameter.\cite{Aji:07,Aji:10} The 
extended model attributes the anomalous normal-state properties of cuprates to a funnel-shaped quantum critical region in the $T$-versus-$p$ phase
diagram that extends to temperatures well above the QCP at $p \! = \! p_c$, $T \! = \! 0$. In the quantum-critical region the CC order spatially and 
temporally fluctuates between four possible ground-state configurations characterized by different directions of the CC order parameter. 
Local disorder is argued to couple to the CC order, leading to four distinct domains 
consisting of one of the four possible CC order configurations. The fluctuation rate between the different CC order configurations has been estimated to be 
slow enough to appear static on the time scale of neutron scattering, but too fast to cause relaxation of $\mu$SR or NMR spectra.\cite{Varma:14} 

One exception to the null local-probe results is a ZF-$\mu$SR study of a large YBa$_2$Cu$_3$O$_{6.6}$ single crystal in which the unusual 3-D IUC 
magnetic order has been detected by polarized neutron scattering.\cite{Sonier:09}  Static magnetic order with an onset temperature and local magnetic 
field consistent with the neutron findings was observed, but only in $\sim \! 3$~\% of the sample. This raises the possibility of fluctuating 
IUC magnetic order (that is not necessarily CC order) being locally pinned in a static configuration by disorder.
The impurity/disorder type must be fairly specific though, since it has been shown that Zn substitution of Cu in YBa$_2$Cu$_3$O$_{6.6}$ 
does not affect the magnetic-onset temperature, but does reduce the magnetic Bragg scattering intensity.\cite{Baledent:11}
In other words, the Zn impurity apparently reduces the volume of the sample containing the IUC magnetic order.

Here we investigate whether there is fluctuating IUC magnetic order that slows down enough near $T \! = \! 0$,
where thermal fluctuations vanish, to become detectable by ZF-$\mu$SR. If the mysterious magnetic order is associated
with a QCP, then near $T \! = \! 0$ we expect quantum fluctuations to dominate close to $p_c$, but in the
absence of significant disorder to have a diminishing effect as the hole concentration is lowered.
The neutron experiments on Y123 and Hg1201 suggest $p_c \! \sim \! 0.19$, and
previous ZF-$\mu$SR measurements on Y-doped Bi2212, pure LSCO, and Zn-doped LSCO, extending down to 40 mK show a vanishing of 
low-frequency spin fluctuations above this critical doping.\cite{Pano:02} However, a similar ZF-$\mu$SR study down to such low temperatures 
has not been performed on the other cuprates in which IUC magnetic order has been detected by neutrons. An exception are ZF-$\mu$SR 
measurements on a $p \! \sim \! 0.167$ Bi2212 powdered sample, which indicate the onset of spin fluctuations below $T \! \sim \! 5$~K, but
no spin freezing down to 40 mK.\cite{Pano:02} 

ZF-$\mu$SR measurements with the initital muon spin polarization {\bf P}(0) parallel to the $\hat{c}$-axis
were performed on underdoped ($p \! = \! 0.094$, $T_c \! = \! 58$~K) and optimally-doped ($p \! = \! 0.16$, $T_c \! = \! 90$~K) 
Bi2212 single crystals, and single crystals of underdoped ($p \! = \! 0.11$, $T_c \! = \! 62.5$~K) YBa$_2$Cu$_3$O$_{6.57}$.
The samples were prepared as described elsewhere.\cite{Lotfi:13,Liang:98}  
Spectra were collected down to as low as $T \! = \! 24$~mK using 
a dilution refrigerator on the M15 surface muon beam line at the TRIUMF subatomic physics laboratory in Vancouver, Canada. The single crystals were
mounted on a silver (Ag) sample holder, covering a 8~mm~$\times$~5~mm area.
A scintillation detector placed downstream was used to reject muons that missed the sample.
A fraction ($\leq \! 40$~\%) of the incoming muons stopped in the uncovered portion of the Ag sample holder, and a fraction ($\sim \! 20$~\%)
of the muons stopped in the copper (Cu) heat shields of the dilution refrigerator. Since the nuclear dipole fields in Ag are negligible, there is 
no appreciable time or temperature dependence to the background component from the sample holder. While the relaxation rate of the
ZF-$\mu$SR signal from Cu does have a temperature dependence caused by muon diffusion,\cite{Kadono:89} the Cu shields are at constant temperature.  
We also performed longitudinal-field (LF) $\mu$SR measurements on $p \! = \! 0.11$ Y123 single crystals at a fixed temperature far 
below $T_c$ using a helium-gas flow cryostat and low-background sample holder, for the purpose of determining whether 
the internal magnetic fields are static or dynamic. In this setup there is no Cu component and the background contribution to the LF-$\mu$SR signal 
is less than 20~\%. 

\begin{figure}
\centering
\includegraphics[width=8.0cm]{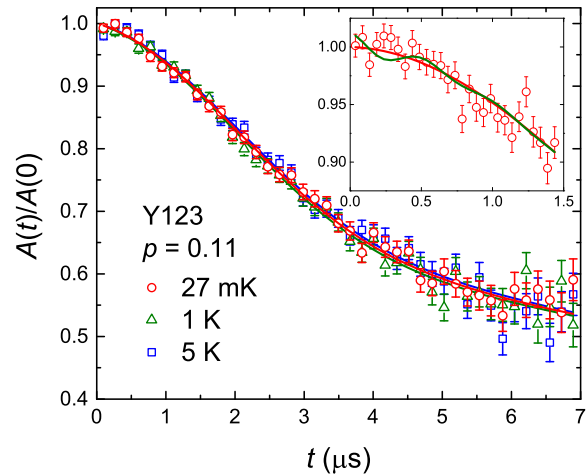}
\caption{(Color online) Representative normalized ZF-$\mu$SR asymmetry spectra for underdoped YBa$_2$Cu$_3$O$_{6.57}$ single crystals. 
These spectra were recorded with the initial muon spin polarization {\bf P}(0) parallel to the $\hat{c}$-axis.  
The solid curves through the data points are fits to Eq.~(\ref{eqn:AsyTot}), assuming Eq.~(\ref{eqn:AsyProduct})
for the relaxation function $G_{\rm s}(t)$. The inset shows the ZF-$\mu$SR spectrum for $T \! = \! 27$~mK at early times.
The solid green curve simulates the presence of a 3~\% damped-oscillating contribution to the sample component
assuming a mean local field of 141~G.}
\label{fig1}
\end{figure}

The ZF-$\mu$SR asymmetry spectra were fit to the sum of sample and backgrounds terms as follows
\begin{equation}
A(t)=a_{\rm s}G_{\rm s}(t) + a_{\rm b}G_{\rm b}(t) \, ,
\label{eqn:AsyTot}
\end{equation}
where $a_{\rm s}$ and $G_{\rm s}(t)$ [$a_{\rm b}$ and $G_{\rm b}(t)$] are the amplitude and ZF
relaxation function for the sample (background) contribution.
The background term originating from muons stopping outside of the sample was assumed to be independent of temperature and 
approximately described by the following relaxation function
\begin{equation}
G_{\rm b}(t) = G_{z}^{\rm KT}(\Delta_{\rm b}, t) \, ,
\end{equation}
where $G_{z}^{\rm KT}(\Delta_{\rm b}, t)$ is a static Gaussian Kubo-Toyabe function. In particular,
\begin{equation}
	  G_{z}^{\rm KT}(\Delta_{\rm b}, t) = \dfrac{1}{3} +\dfrac{2}{3}(1-\Delta_{\rm b}^2 t^2) \exp\left(-\dfrac{1}{2}\Delta_{\rm b}^2 t^2 \right) \, ,
\end{equation}
where $\gamma_\mu$ is the muon gyromagnetic ratio and ${\Delta_{\rm b}}/{\gamma_\mu}$ is the width of the Gaussian distribution in field 
sensed by the implanted muon ensemble. The sample contribution was assumed to be the product of two relaxation functions  
\begin{equation}
G_{\rm s}(t) = G_{z}^{\rm KT}(\Delta_{\rm s}, t) \exp(-\lambda t) \, ,
\label{eqn:AsyProduct}
\end{equation}
which assumes that muons stopping in the sample sense the vector sum of static nuclear dipolar fields and
fields of some other origin that generate a weak exponential relaxation rate $\lambda$.
An exception is Bi2212 at $p \! = \! 0.094$, where the ZF-$\mu$SR asymmetry spectra below $T \! = \! 1$~K
were better described by
\begin{equation}
G_{\rm s}(t) = [f \exp(-\eta t) + (1-f)]G_{z}^{\rm KT}(\Delta_{\rm s}, t) \exp(-\lambda t) \, .
\label{eqn:AsySum}
\end{equation}
This function assumes an enhanced exponential relaxation rate $\lambda \! + \! \eta$ due to a fraction $f$ of the 
muons experiencing additional fields in some parts of the sample.
In contrast to the relaxation rates $\Delta_{\rm s}$ and $\Delta_{\rm b}$, the exponential relaxation rates $\lambda$ and $\eta$
were allowed to vary with temperature in the fitting of the ZF-$\mu$SR spectra. In addition, $f$ was assumed to be
independent of temperature.

\begin{figure}
\centering
\includegraphics[width=8.0cm]{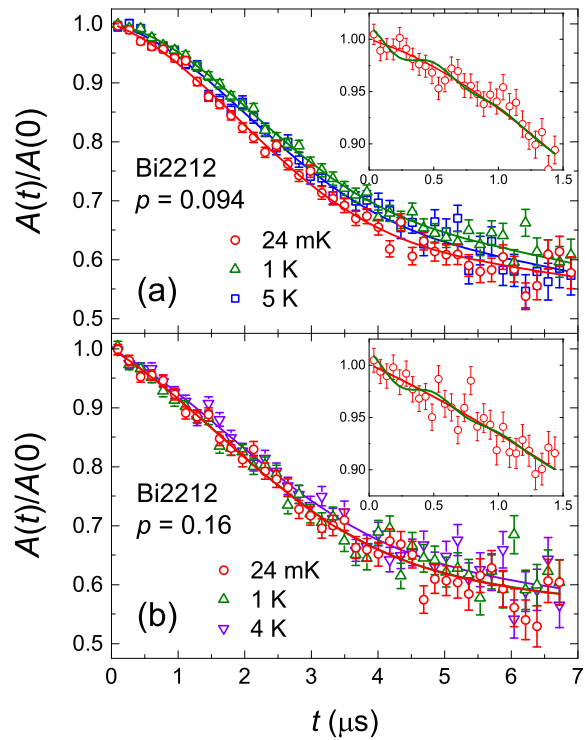}
\caption{(Color online) Representative normalized ZF-$\mu$SR asymmetry spectra for (a) underdoped and (b) optimally-doped 
Bi2212 single crystals. The insets show the ZF-$\mu$SR spectra for the lowest temperature at early times.
These spectra were recorded with the initial muon spin polarization parallel to the $\hat{c}$-axis.    
The solid curves through the data points are fits to Eq.~(\ref{eqn:AsyTot}), assuming Eq.~(\ref{eqn:AsyProduct})
for the relaxation function $G_{\rm s}(t)$. An exception is the solid curve for the $p \! = \! 0.094$ sample at 
$T \! = \! 24$~mK, which is a fit assuming Eq.~(\ref{eqn:AsySum}) for $G_{\rm s}(t)$.
The insets show the ZF-$\mu$SR spectra for $T \! = \! 24$~mK at early times.
The solid green curves simulate the presence of a 3~\% damped-oscillating contribution to the sample component
assuming a mean local field of 141~G.}
\label{fig2}
\end{figure}

\begin{figure}
\centering
\includegraphics[width=8.5cm]{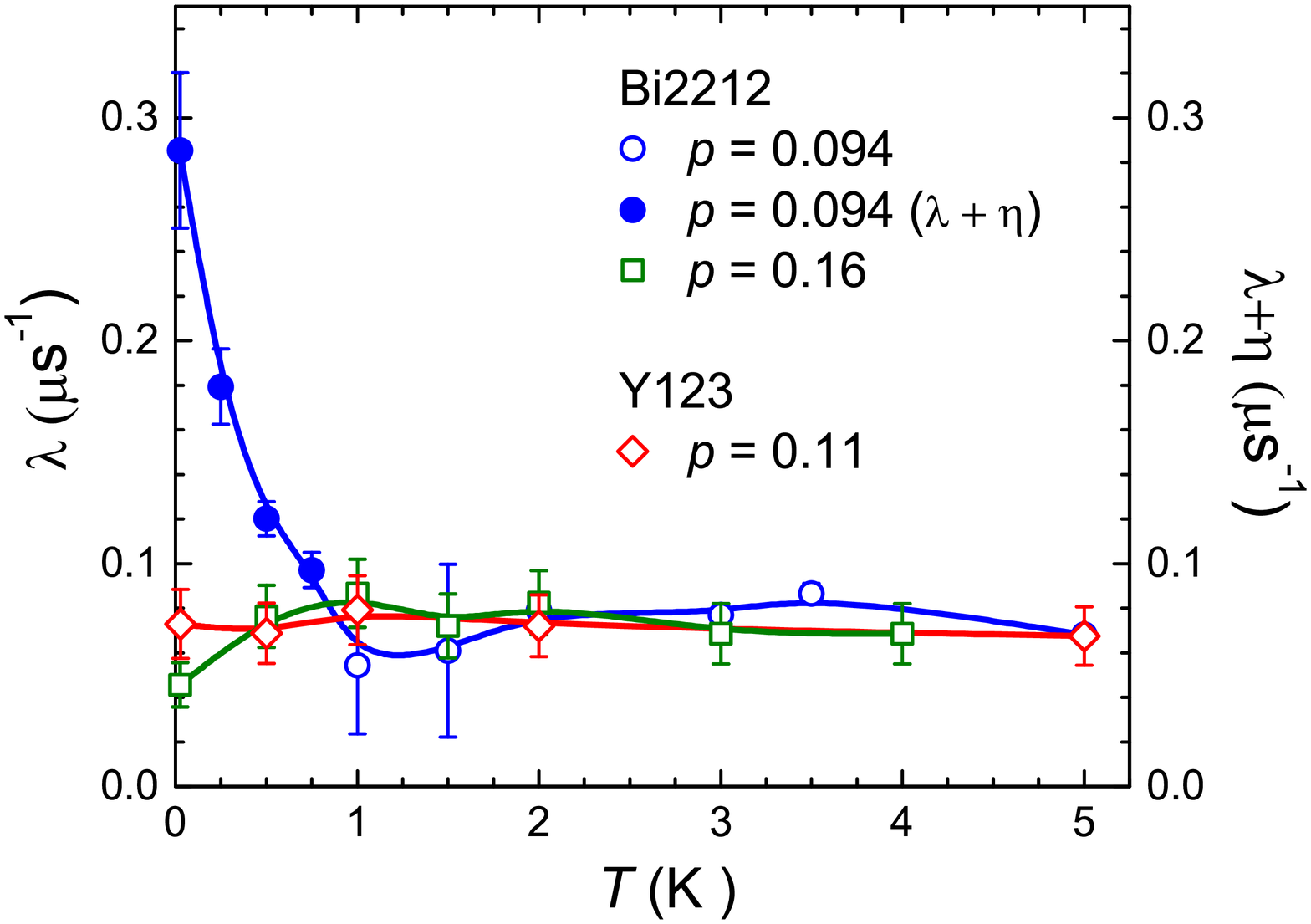}
\caption{(Color online) Temperature dependence of the ZF exponential relaxation rate $\lambda$ (open symbols).
Also shown is the enhanced exponential relaxation rate $\lambda \! + \! \eta$ (solid circles) 
for $p \! = \! 0.094$ Bi-2212 below $T \! = \! 1$~K, which is due to a fraction (34.6~\%) of the implanted 
muons experiencing additional internal magnetic fields.}
\label{fig3}
\end{figure}

Figures~\ref{fig1} and \ref{fig2} show representative ZF-$\mu$SR asymmetry spectra for the Y123 and 
Bi2212 samples. The fits described above yielded $\Delta_{\rm s} \! = \! 0.156(1)$~$\mu$s$^{-1}$ and $\Delta_{\rm b} \! = \! 0.405(6)$~$\mu$s$^{-1}$ 
for the $p \! = \! 0.11$ Y123 sample, and
$\Delta_{\rm s} \! = \! 0.134(6)$~$\mu$s$^{-1}$ and $\Delta_{\rm b} \! = \! 0.395(1)$~$\mu$s$^{-1}$
($\Delta_{\rm s} \! = \! 0.134(6)$~$\mu$s$^{-1}$ and $\Delta_{\rm b} \! = \! 0.393(6)$~$\mu$s$^{-1}$) for the
$p \! = \! 0.094$ ($p \! = \! 0.16$) Bi2212 sample. The values of $\Delta_{\rm b}$ are consistent with the
background relaxation being dominated by the Cu heat shields.   
Below $T \! = \! 1$~K the fits of the $p \! = \! 0.094$ Bi2212 ZF-$\mu$SR
signals yielded $f \! = \! 0.346$, indicating that about one third of the muons implanted in the sample
sense local magnetic fields in addition to the host magnetic nuclear dipole moments. The ZF-$\mu$SR spectra do
not exhibit an oscillatory component indicative of long-range magnetic order. The presence of a magnetically-ordered
state with a broad distribution of local magnetic fields or a small magnetically-ordered volume fraction 
would result in a rapidly damped oscillatory signal. The insets of Figs.~\ref{fig1} and \ref{fig2} show
the ZF-$\mu$SR signal for $T \! \leq \! 27$~mK plotted over the first 1.5~$\mu$s. While there is no apparent
oscillatory component, simulations of a 3~\% magnetically-ordered phase of the kind observed in the large
YBa$_2$Cu$_3$O$_{6.60}$ single crystal in Ref.~\onlinecite{Sonier:09} superimposed on the early-time
ZF-$\mu$SR spectra (green curves in the insets of Figs.~\ref{fig1} and \ref{fig2}) show that a small
$0.4 \! \times \ 3$~\%~$\! = \! 1.2$~\% contribution to the total signal cannot be ruled out. However, it
is worth mentioning that no such minority phase was previously observed in low-background measurements of the $p \! = \! 0.11$ Y123 sample 
above $T \! = \! 2.3$~K.\cite{Sonier:09}  

\begin{figure}
\centering
\includegraphics[width=8.5cm]{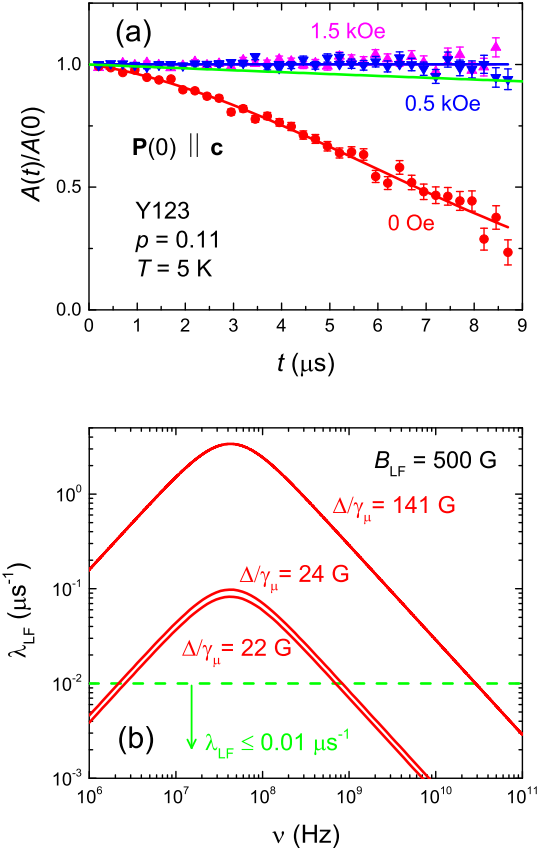}
\caption{(Color online) (a) Normalized LF-$\mu$SR asymmetry spectra for underdoped YBa$_2$Cu$_3$O$_{6.57}$ single 
crystals at $T \! = \! 5$~K, recorded with the initial muon spin polarization {\bf P}(0) parallel to the $\hat{c}$-axis. The circles, nablas,
and triangles correspond to data recorded in external longitudinal magnetic fields of $H \! = \! 0$, 0.5 and 1.5~kOe, respectively.
The green curve is the sum of a pure exponential function and a constant, $0.8\exp(-\lambda_{\rm LF} t) \! + \! 0.2$, where 
$\lambda_{\rm LF} \! = \! 0.01$~$\mu$s$^{-1}$.    
(b) Calculated relaxation rate $\lambda_{\rm LF}$ from Eq.~(\ref{eqn:Redfield}) for a longitudinal magnetic field of $0.5$~kOe and
different values of the local-field fluctuation amplitude $\Delta/\gamma_{\mu}$. 
The dashed green line indicates the maximum value of $\lambda_{\rm LF}$ inferred from the corresponding LF-$\mu$SR spectrum in (a).}
\label{fig4}
\end{figure}

Figure~\ref{fig3} shows the temperature dependence of the exponential relaxation rate $\lambda$ for all
three samples, along with $\lambda \! + \! \eta$ for $p \! = \! 0.094$ Bi2212 below $T \! = \! 1$~K.
While there is an increase in the relaxation rate for the $p \! = \! 0.094$ Bi2212 sample
below $T \! = \! 1$~K, this is most likely due to low-energy spin fluctuations in the CuO$_2$ planes,
as spin freezing is observed in Y-doped Bi2212 below $p \! \sim \! 0.10$.\cite{Pano:02} 
The lack of any increase of $\lambda$ at low temperatures for the $p \! = \! 0.11$ Y123 and the 
$p \! = \! 0.16$ Bi2212 samples rules out the onset of quasi-static magnetism 
below $T \! = \! 5$~K. However, these ZF-$\mu$SR results do not rule out the possibility that even at these low 
temperatures and at a hole doping far below $p_c$, the IUC magnetic order fluctuates too fast to be 
detectable on the time scale of ZF-$\mu$SR. Assuming the local magnetic field due 
to IUC magnetic order is 141~G (as estimated in Ref.~\onlinecite{Sonier:09}), the ZF-$\mu$SR results 
for $p \! = \! 0.11$ Y123 and $p \! = \! 0.16$ Bi2212 imply a lower limit of $1.9 \! \times \! 10^6$~Hz
for the fluctuation rate. This is far below the upper limit of $10^{11}$~Hz imposed by the 
energy resolution of the polarized neutron experiments.

Our LF-$\mu$SR measurements in a different experimental setup greatly increase the lower limit of 
the fluctuation rate. Figure~\ref{fig4}(a) shows LF-$\mu$SR spectra recorded for $p \! = \! 0.11$ Y123 well below $T_c$.   
Below $T_c$ weak applied fields are completely or partially screened from the bulk, and hence external fields 
well in excess of the lower critical field $H_{c1}$ were applied. A longitudinal field of $B_{\rm LF} \! = \! 0.5$~kOe 
completely decouples the muon spin from the nuclear dipoles of the background and the internal magnetic fields of the sample. 
If the muons sense a rapidly fluctuating nearly Gaussian distribution of field, the ZF-$\mu$SR signal will decay 
with a pure exponential relaxation $G_{\rm s}(t) \! = \! \exp(- \lambda t)$. In this case the dependence of the dynamic relaxation rate on the 
LF is given by the Redfield equation\cite{Slichter:90} 
\begin{equation}
\lambda_{\rm LF} = \frac{2 \Delta^2/\nu}{1 + (\gamma_{\mu} B_{\rm LF}/\nu)^2}  \, ,
\label{eqn:Redfield}
\end{equation}
where $\Delta/\gamma_{\mu}$ is the width of the field distribution and $\nu$ is the local-field fluctuation frequency.
In the previous $\mu$SR study of a large single crystal of YBa$_2$Cu$_3$O$_{6.6}$ in which static magnetic order
was detected in 3~\% of the sample,\cite{Sonier:09} the mean local field detected was $\sim \! 141$~G --- which was
shown to be in good agreement with the magnitude and direction of the ordered moment determined by polarized neutron
diffraction. Figure~\ref{fig4}(b) shows a simulation of the dependence of $\lambda_{\rm LF}$ on $\nu$ for 
a LF of $B_{\rm LF} \! = \! 500$~G and different values of the local-field fluctuation amplitude $\Delta/\gamma_{\mu}$.
The values $\Delta/\gamma_{\mu} \! = \! 141$~G and 24~G assume the polarized neutron measurements of $p \! = \! 0.11$ Y123 
(Ref.~\onlinecite{Mook:08}) detect IUC magnetic order within the CuO$_2$ planes in 3~\% and 100~\% of the sample, respectively.
Also shown is the upper limit $\lambda_{\rm LF} \! \leq \! 0.01$~$\mu$s$^{-1}$ inferred from the corresponding LF-$\mu$SR spectrum in Fig.~\ref{fig4}(a)
under the assumption that fluctuating magnetism occurs throughout the sample. 
The simulation of $\lambda_{\rm LF}$ for $\Delta/\gamma_{\mu} \! = \! 141$~G exceeds the upper limit of the relaxation rate 
observed in $p \! = \! 0.11$ Y123 below $\nu \! \sim \! 3 \! \times \! 10^{10}$~Hz. On the other hand, the simulation for 
$\Delta/\gamma_{\mu} \! = \! 24$~G only rules out a fluctuation rate below $\nu \! \sim \! 10^{9}$~Hz.
If the IUC magnetic order is due to loop currents flowing out of the CuO$_2$ plane through the apical oxygen as
proposed in Ref.~\onlinecite{Weber:09}, $\Delta/\gamma_{\mu} \! = \! 22$~G and the lower limit of the fluctuation rate
is slightly reduced. Regardless, the combined LF-$\mu$SR results and the polarized neutron measurements 
place narrow limits of $10^{9}$ to $10^{11}$~Hz on the fluctuation rate of the IUC magnetic order.            

If the IUC magnetic order is associated with fluctuations between different orientations of a CC-ordered 
state in finite size domains, rather than spatially-uniform long range magnetic order, quantum fluctuations will 
not diminish away from the QCP.\cite{Varma:14} The lowest quantum fluctuation frequency between the distinct CC 
configurations is estimated to be less than 10$^{10}$~Hz --- a scenario not completely ruled out by our LF-$\mu$SR results.
As for other possible origins of the IUC magnetic order, while our estimated lower limit 
of the fluctuation frequency assumes fluctuating magnetic order throughout the sample volume, the current
measurements do not rule out the possibility that there is slower fluctuating IUC magnetic order contained
in a small volume fraction.

\begin{acknowledgments}
We thank the staff of TRIUMF's Centre for Molecular and Materials Science for technical assistance. 
JES, WNH, DAB, and RL acknowledge support from CIFAR and NSERC of Canada.
\end{acknowledgments}


\begin{thebibliography}{10}

\bibitem{Fauque:06} B. Fauqu\'{e}, Y. Sidis, V. Hinkov, S. Pailh\`{e}s, C. T. Lin, X. Chaud, and P. Bourges, Phys. Rev. Lett. {\bf 96}, 197001 (2006). 

\bibitem{Mook:08} H. A. Mook, Y. Sidis, B. Fauqu\'{e}, V. Bal\'{e}dent, and P. Bourges, Phys. Rev. B {\bf 78}, 020506(R) (2008).

\bibitem{Li:08} Y. Li, V. Bal\'{e}dent, N. Bari\u{s}i\'{c}, Y. Cho, B. Fauqe\'{e}, Y. Sidis, G. Yu, X. Zhao, P. Bourges, and M. Greven, Nature {\bf 455}, 372 (2008).

\bibitem{Baledent:11} V. Bal\'{e}dent, D. Haug, Y. Sidis, V. Hinkov, C. T. Lin and P. Bourges, Phys. Rev. B {\bf 83}, 104504 (2011).

\bibitem{Li:11} Y. Li, V. Bal\'{e}dent, N. Bari\u{s}i\'{c}, Y. C. Cho, Y. Sidis, G. Yu, X. Zhao, P. Bourges, and M. Greven, Phys. Rev. B {\bf 84}, 224508 (2011).

\bibitem{Didry:12} S. De Almeida-Didry, Y. Sidis, V. Bal\'{e}dent, F. Giovannelli, I. Monot-Laffez, and P. Bourges Phys. Rev. B {\bf 86}, 020504(R) (2012).

\bibitem{Thro:14} L. Mangin-Thro, Y. Sidis, P. Bourges, S. De Almeida-Didry, F. Giovannelli, and I. Laffez-Monot, Phys. Rev. B {\bf 89}, 094523 (2014).

\bibitem{Thro:15} L. Mangin-Thro, Y. Sidis, A. Wildes, and P. Bourges, Nat. Commun. {\bf 6}, 7705 (2015).

\bibitem{Xia:08} J. Xia, E. Schemm, G. Deutscher, S. A. Kivelson, D. A. Bonn, W. N. Hardy, R. Liang, W. Siemons, G. Koster, M. M. Fejer, 
and A. Kapitulnik, Phys. Rev. Lett. {\bf 100}, 127002 (2008).

\bibitem{Leridon:09} B. Leridon, P. Monod, D. Colson, and A. Forget, Europhys. Lett. {\bf 87}, 17011 (2009).

\bibitem{He:11} R.-H. He, M. Hashimoto, H. Karapetyan, J. D. Koralek, J. P. Hinton, J. P. Testaud, V. Nathan, Y. Yoshida, H. Yao, K. Tanaka, W. Meevasana, 
R. G. Moore, D. H. Lu, S.-K. Mo, M. Ishikado, H. Eisaki, Z. Hussain, T. P. Devereaux, S. A. Kivelson, J. Orenstein, A. Kapitulnik, and Z.-X. Shen, 
Science {\bf 331}, 1579-1583 (2011).

\bibitem{Shekhter:13} A. Shekhter, B. J. Ramshaw, R. Liang, W. N. Hardy, D. A. Bonn, F. F. Balakirev, R. D. McDonald, J. B. Betts, S. C. Riggs,
and A. Migliori, Nature {\bf 498}, 75-77 (2013).  

\bibitem{Baledent:10} V. Bal\'{e}dent, B. Fauqu\'{e}, Y. Sidis, N.B. Christensen, S. Pailh\`{e}s, K. Conder, E. Pomjakushina, J. Mesot, and P. Bourges, 
Phys. Rev. Lett. {\bf 105}, 027004 (2010).

\bibitem{Varma:97} C.M. Varma, Phys. Rev. B {\bf 55}, 14554 (1997).

\bibitem{Sonier:02} J.E. Sonier, J. H. Brewer, R. F. Kiefl, R. H. Heffner, K. F. Poon, S. L. Stubbs, G. D. Morris, R. I. Miller, W. N. Hardy, R. Liang, 
D. A. Bonn, J. S. Gardner, C. E. Stronach, and N. J. Curro, Phys. Rev. B {\bf 66}, 134501 (2002).

\bibitem{MacDougall:08} G. J. MacDougall, A. A. Aczel, J. P. Carlo, T. Ito, J. Rodriguez, P. L. Russo, Y. J. Uemura, S. Wakimoto, and G. M. Luke, 
Phys. Rev. Lett. {\bf 101}, 017001 (2008).

\bibitem{Sonier:09} J. E. Sonier, V. Pacradouni, S. A. Sabok-Sayr, W. N. Hardy, D. A. Bonn, R. Liang, and H. A. Mook, Phys. Rev. Lett. {\bf 103}, 167002 (2009).

\bibitem{Huang:12} W. Huang, V. Pacradouni, M. P. Kennett, S. Komiya, and J. E. Sonier, Phys. Rev. B {\bf 85}, 104527 (2012).

\bibitem{Shekhter:08} A. Shekhter, L. Shu, V. Aji, D. E. MacLaughlin and C. M. Varma, Phys. Rev. Lett. {\bf 101}, 227004 (2008).

\bibitem{Kaiser:12} C.V. Kaiser, W. Huang, S. Komiya, N.E. Hussey, T. Adachi, Y. Tanabe, Y. Koike, and J. E. Sonier, Phys. Rev. B {\bf 86}, 054522 (2012).

\bibitem{Sonier:13} J.E. Sonier, W Huang, V. Pacradouni, M.P. Kennett, and S. Komiya, J. Phys.: Conf. Ser. {\bf 449}, 012013 (2013).

\bibitem{Strassle:08} S. Str\"{a}ssle, J. Roos, M. Mali, H. Keller, and T. Ohno, Phys. Rev. Lett. {\bf 101}, 237001 (2008).

\bibitem{Strassle:10} S. Str\"{a}ssle, B. Graneli, M. Mali, J. Roos, and H. Keller, Phys. Rev. Lett. {\bf 106}, 097003 (2011).

\bibitem{Mounce:13} A. M. Mounce, Sangwon Oh, Jeongseop A. Lee, W. P. Halperin, A. P. Reyes, P. L. Kuhns, M. K. Chan, C. Dorow, L. Ji, D. Xia, X. Zhao, 
and M. Greven, Phys. Rev. Lett. {\bf 111}, 187003 (2013).

\bibitem{Wu:15} T. Wu, H. Mayaffre, S. Kr\"{a}mer, M. Horvati\'{c}, C. Berthier, W.N. Hardy, R. Liang, D.A. Bonn, and M.-H. Julien, Nature Commun. {\bf 6}, 6438 (2015).

\bibitem{Kung:14} Y. F. Kung, C.-C. Chen, B. Moritz, S. Johnston, R. Thomale, and T. P. Devereaux, Phys. Rev. B {\bf 90}, 224507 (2014).

\bibitem{Moskvin:12} A. S. Moskvin, JETP Lett. {\bf 96}, 385 (2012).

\bibitem{Aji:07} V. Aji and C.M. Varma, Phys. Rev. Lett. {\bf 99}, 067003 (2007).

\bibitem{Aji:10} V. Aji, A. Shekhter, and C. M. Varma, Phys. Rev. B {\bf 81}, 064515 (2010).

\bibitem{Varma:14} C. M. Varma, J. Phys.: Condens. Matter {\bf 26}, 505701 (2014).

\bibitem{Pano:02} C. Panagopoulos, J. L. Tallon, B. D. Rainford, T. Xiang, J. R. Cooper, and C. A. Scott, Phys. Rev. B {\bf 66}, 064501 (2002). 

\bibitem{Lotfi:13} Z. L. Mahyari, A. Cannell, E. V. L. de Mello, M. Ishikado, H. Eisaki, R. Liang, D. A. Bonn, and J.E. Sonier,
Phys. Rev. B {\bf 88}, 144504 (2013).

\bibitem{Liang:98} R. Liang, D. A. Bonn, and W. N. Hardy, Physica C {\bf 304}, 105 (1998).

\bibitem{Kadono:89} R. Kadono, J. Imazato, T. Matsuzaki, K. Nishiyama, K. Nagamine, and T. Yamazaki, Phys. Rev. B {\bf 39}, 23 (1989).

\bibitem{Slichter:90} C. P. Slichter, {\it Principles of Magnetic Resonance}, 3rd ed. (Springer-Verlag, Berlin, 1990).

\bibitem{Weber:09} C. Weber, A. L\"{a}uchli, F. Mila and T. Giamarchi, Phys. Rev. Lett. {\bf 102}, 017005 (2009). 
 
\end{thebibliography}
\end{document}